\documentclass[12pt]{article}
\usepackage{a4wide}

\def\Name#1{#1}
\def\And{{and }}
\def\Review#1{{\em #1}}
\def\Vol#1{{\bf #1}}
\def\Year#1{#1}
\def\Page#1{#1}
\newfont{\QRC}{msbm10 scaled\magstep1}
\def\ev#1{\left<#1\right>}
\def\av#1{\left|#1\right|}
\def\pa#1{\left(#1\right)}

\usepackage{graphics}
\DeclareGraphicsExtensions{ps,eps,epsf}

\def\myfig#1#2#3{
  \begin{figure}
    \centerline{
    \resizebox{0.7\textwidth}{!}{\includegraphics{#1}}}
    \vspace{-14pt}
    \caption{#2}
    \label{#3}
  \end{figure}
}

\def\myfigSS#1#2#3{
  \begin{figure}
    \centerline{
    \resizebox{0.8\textwidth}{!}{\includegraphics{#1}}}
    \caption{#2}
    \label{#3}
  \end{figure}
}

\newcommand{\figCurvature}{ 
  \myfig{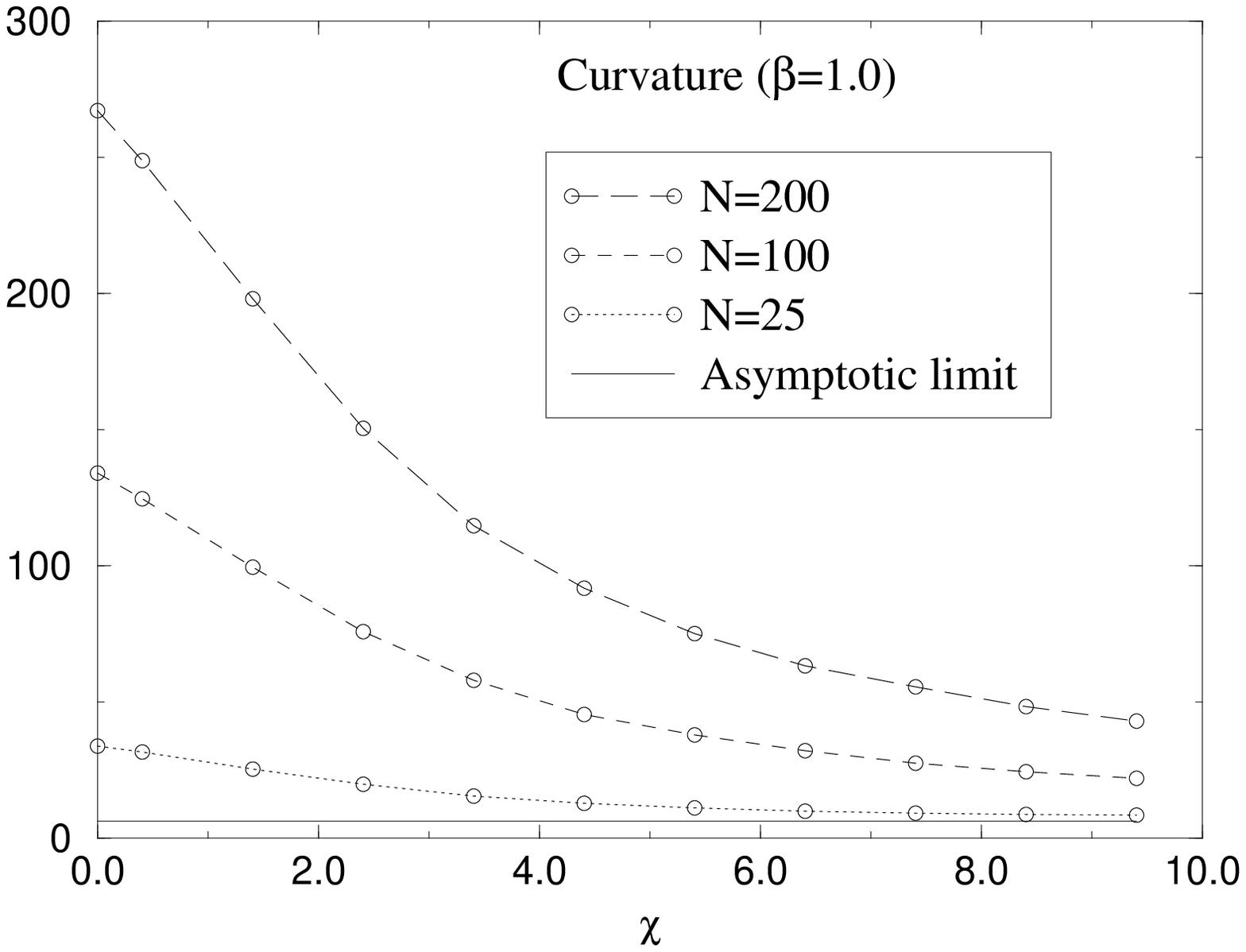}{Curvature for several paths}{fig:Curvature}}

\newcommand{\figHausdorff}{ 
  \myfig{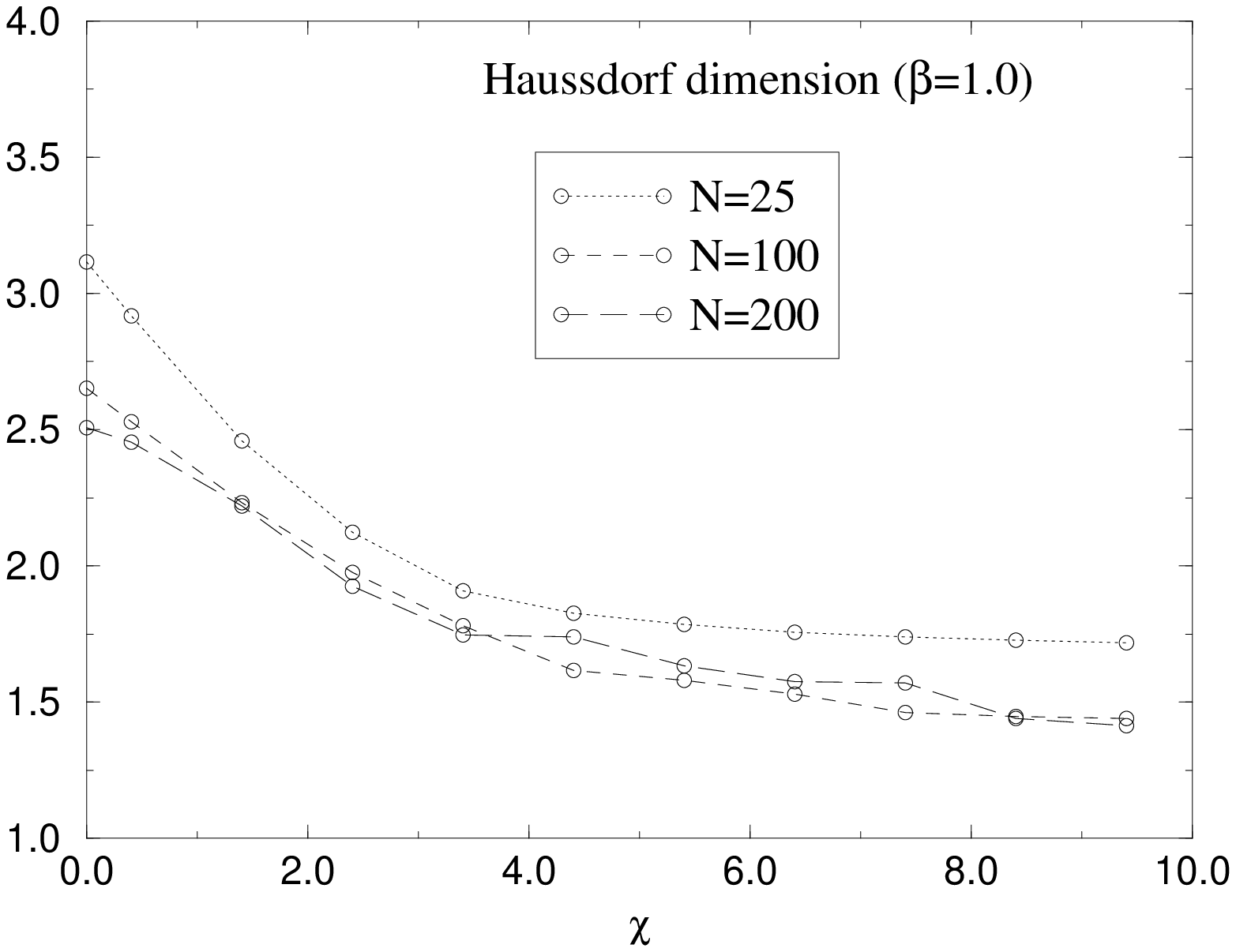}{Hausdorff dimension for several paths}{fig:Hausdorff}}

\newcommand{\figCv}{ 
  \myfig{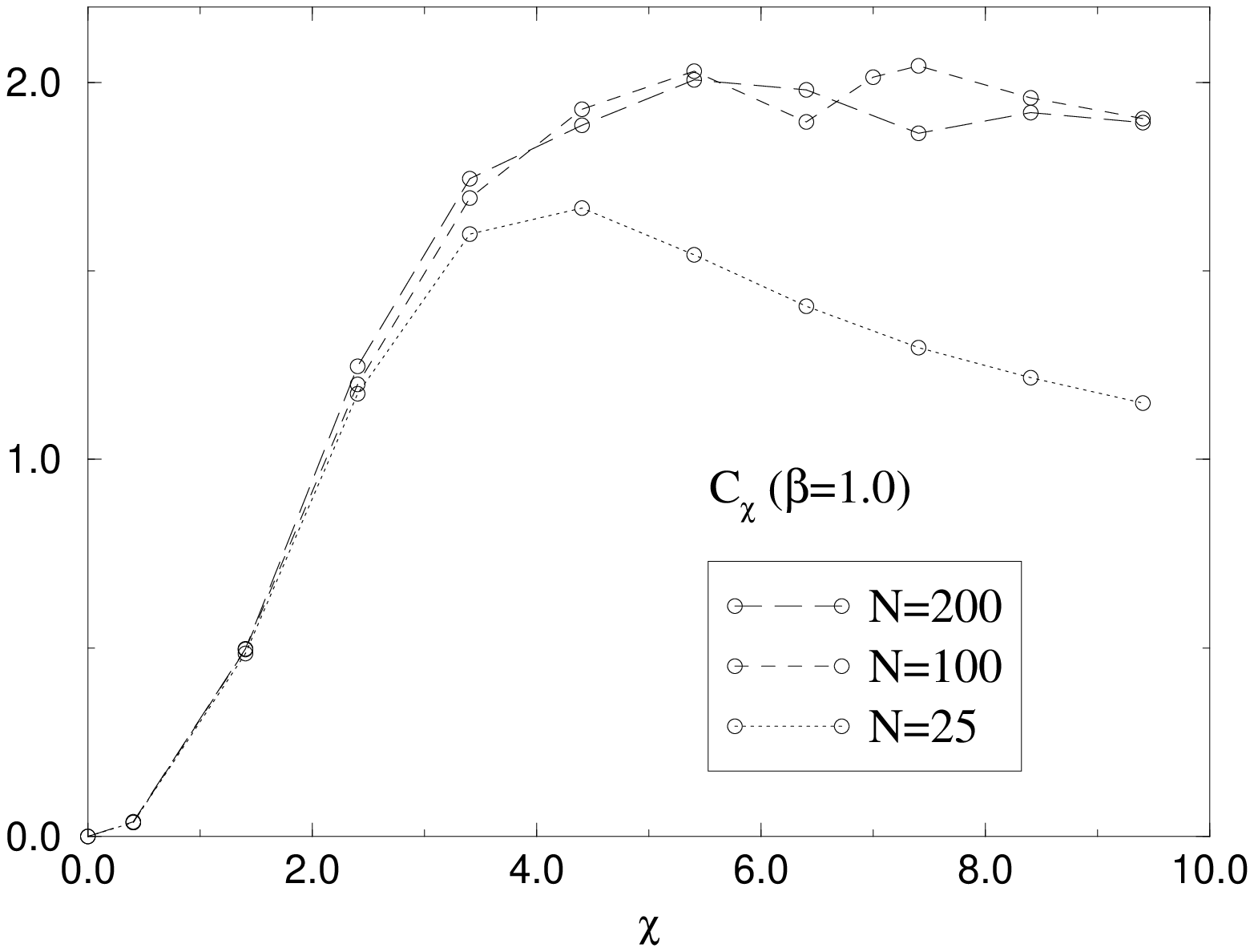}{Specific heat for several paths.}{fig:Cv}}

\newcommand{\figSS}{ 
  \myfigSS{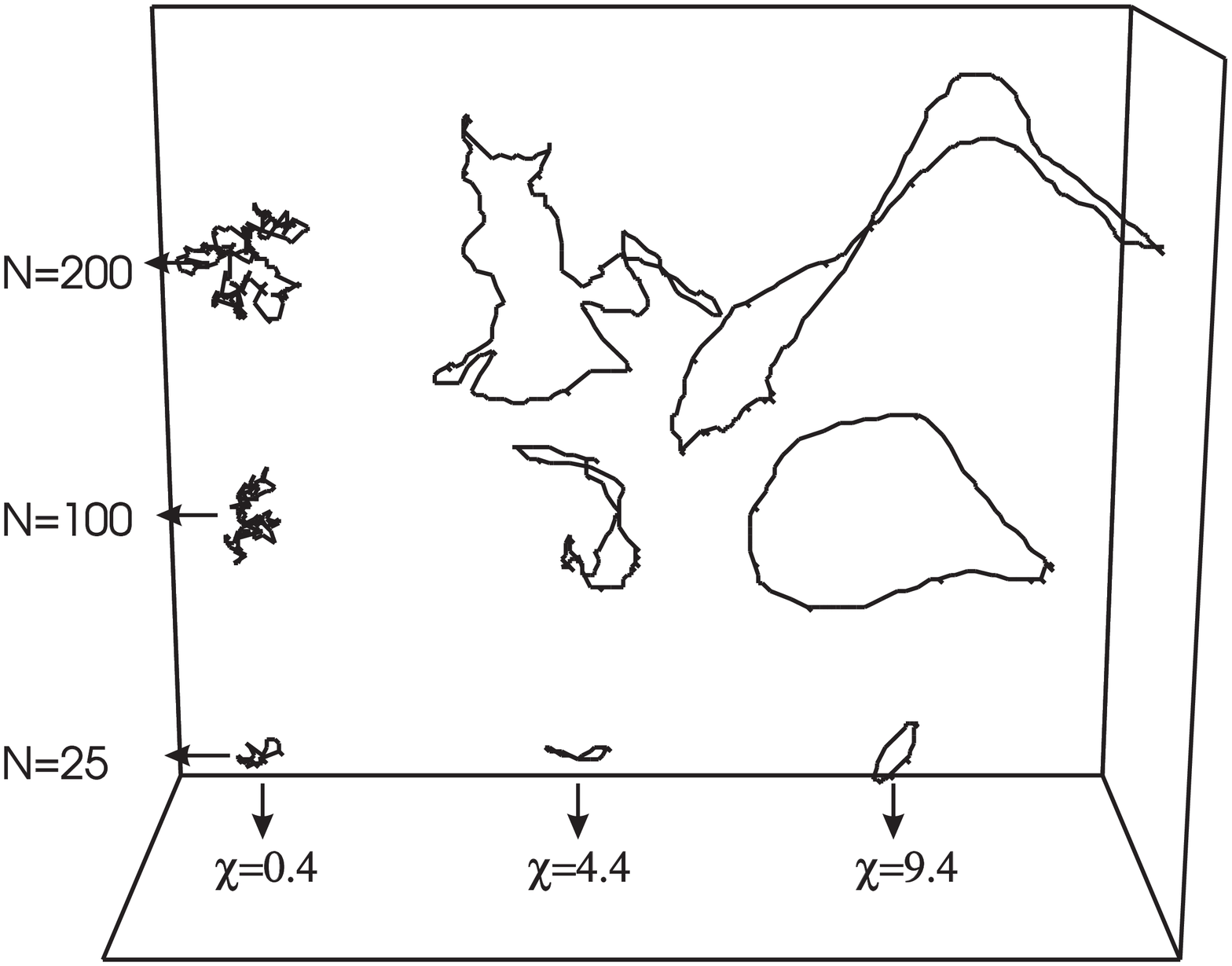}{Simple snapshots for several paths.}{fig:SS}}

\newcommand{\tabLenghts}{
\begin{table}[hb]
\caption{Some lengths compared with the predicted value}
$$
\begin{array}{|c|c|c|c|}
\hline
\chi & N=25 \ (72) & N=100 \ (297)& N=200 \ (597) \\ \hline
0.0  & 71.9\pm 0.1 & 296.7\pm 0.2 & 597.3\pm 0.4  \\
1.4  & 71.7\pm 0.1 & 296.5\pm 0.2 & 595.9\pm 0.5  \\
3.4  & 72.3\pm 0.2 & 297.5\pm 0.6 & 600.1\pm 1.0  \\
5.4  & 71.6\pm 0.2 & 297.5\pm 0.8 & 594.8\pm 1.6  \\ \hline
\end{array}
$$
\label{tab:Lenghts}
\end{table}
}

\begin{document}

\title{\bf Numerical simulation of random paths \\
  with a curvature dependent action}
\author{
  {\sc M. Baig, J. Clua} \\
  Grup de F{\'\i}sica Te{\`o}rica, IFAE.
  Universitat Aut{\`o}noma de Barcelona\\
  E-08193 Bellaterra (Barcelona), Spain.\\
  \and\\
  {\sc A. Jaramillo}\\
  Dep. de F{\'\i}sica Te{\`o}rica and I.F.I.C.\\
  Centre Mixt Univ. de Val{\`e}ncia -- C.S.I.C.\\
  E-46100 Burjassot (Val{\`e}ncia), Spain.\\
}

\renewcommand{\today}{}

\maketitle

\begin{abstract}
We study an ensemble of closed random paths, embedded in $\mbox{\QRC R}^3$,
with a curvature dependent action. 
Previous analytical results indicate that
there is no crumpling transition for any finite value of the curvature 
coupling. Nevertheless, in a high statistics  numerical simulation, we observe
two different regimes for the specific heat separated by a
rather smooth structure. The analysis of this fact
warns us about the difficulties in the interpretation of numerical
results obtained in cases where theoretical results are absent and a high
statistics simulation is unreachable. This may be the case of random surfaces.
\end{abstract}

\section{Introduction}
It is well known that the addition of an extrinsic curvature term to the
Nambu-Goto action ({\it i.e.\/} the area action) for the case of random
surfaces controls the formation of {\em spikes}, {\it i.e.\/} deformations
that are exact zero-modes of the area action and causes the degeneration of
the random surface into branched polymers~ \cite{PL85,POL86,DAV86}.

The presence of a second order phase transition for a finite value of the
curvature-coupling separating a {\em crumpled} and a {\em flat} phase for the
case of fixed connectivity random surfaces in three dimensions has been firmly
established by recent large scale numerical simulations~ \cite{WHE96,BCFTA96}.

Nevertheless, the situation is far from being so clear for the case of
dynamically triangulated random surfaces with extrinsic curvature (DTRS).
Results from numerical simulations of this model are unclear and even the real
existence of a phase transition separating the two regimes has been
questioned~ \cite{BCHHM93,ABCFHHM93}.

It has been pointed out also that these approaches, initially developed to
give a lattice discretization for the string theories~\cite{A94}, may have a
realization in condensed matter physics as models of membranes or interfaces.
In this sense, fixed connectivity random surfaces are called {\em crystalline}
surfaces and dynamically triangulated surfaces are called {\em fluid}
surfaces.  The comparison between the behavior of DTRS and the {\em true}
fluid membranes states an in\-te\-res\-ting question.  In fact, the Monte
Carlo analysis of the scaling properties of self-avoiding fluid membranes with
an extrinsic bending rigidity indicates that fluid membranes are always
crumpled at sufficiently long length scales~\cite{KG92a,KG92b}.  The observed
peak in the specific heat is interpreted as the effect originated when the
persistence length describing the exponential decay of the normal-normal two
point function in the crumpled regime simply reaches the finite size of the
system at the apparent transition. Can this behavior be translated to the case
of DTRS?

In this letter we propose a simple scenario: the numerical simulation of the
behavior of a set of random paths with an extrinsic curvature term in the
action.  This system is is particularly adequate to test the numerical
techniques used in random surfaces for several reasons:
\begin{enumerate}
\item Its numerical simplicity (not triviality!) that allows to perform
  simulations with several millions of sweeps (simply impossible to do in
  random surfaces).
\item It is the most natural place to start with random surfaces simulations,
  as it has been pointed out by several authors~\cite{BGM84,PIS86,AE87}.
\item It exist a solid framework given by the analytical results obtained by
 Ambj\o rn   et al.~\cite{ADJ87}.
\end{enumerate}

\section{Random Paths with Curvature}
\label{sec:RandomPaths}
The analysis of a random path consisting on straight segments embedded in
$\mbox{\QRC R}^3$ was performed as a one-dimensional analogue to check the
numerical techniques applied in the numerical simulations of triangulated
random surfaces~\cite{BGM84}.  Pisarski~\cite{PIS86} proposed the addition of
an extra term --proportional to the curvature-- to the theory of random paths,
claiming that this theory might be relevant to polymer physics. He noticed the
observation of asymptotic freedom and pointed out the similarities between
this theory and a nonlinear $\sigma$-model with long-range interactions.
Furthermore, Alonso and Espriu~\cite{AE87} included random paths as a simple
version of the random surface theory in a mean field analysis.  They concluded
that the Hausdorff dimension of the paths will change from $d_H=2$ (Brownian
paths) to $d_H=1$ at infinite curvature coupling.  In 1987 Ambj\o rn, Durhuus
and J{\'o}hnsson~\cite{ADJ87} performed a complete analytical analysis of this
model and they concluded that for finite values of the coupling constant of
the curvature term, the theory belongs to the same universality class of
simple random walks.  A non-trivial scaling limit is only possible if the
curvature coupling tends to infinity. They computed exactly the two-point
function in this limit. They observed that all the paths have anomalous
dimension $\eta=1$, which is consistent with the mean field analysis.
 
In this letter we perform an actual simulation of a micro-canonical set of
random paths with a curvature dependent action.  Such an analysis, dealing
with finite paths made of a given number of straight segments, allows for a
graphical visualization of the role of the curvature term of the action.
Geometrical magnitudes such as the curvature or the gyration radius can be
directly measured and the scaling behavior explicitly deduced.  A close
correspondence with the conventions and techniques used in the simulations of
random surfaces has been imposed: closed paths of a given number of points,
fixed topology, a Metropolis algorithm, etc.

\section{Lattice action for closed paths}
Pisarski~\cite{PIS86} considered a one dimensional analogue of the string
action defined in~ \cite{POL86} allowing a path integration over paths
$X^\mu(\psi)$ as
\begin{equation}
S = \beta \int ds + \chi \int k \,\, ds ,
\end{equation}
where $s$ is the parameter for which $\left(\frac{dX}{ds}\right)^2 = 1$, and
$k$ is the curvature given by $k^2=\left( \frac{d^2X}{ds^2}\right)^2$ (both
$\beta$ and $\chi$ being positive).\footnote{ As pointed out in~\cite{ADJ87},
  the curvature term must be proportional to $k$, which is scale invariant,
  and requires a dimensionless coupling.  Terms proportional to higher powers
  are formally irrelevant in the continuum limit.}

Given an arbitrary parameterization
\begin{equation}
S = \chi \int \frac{(\dot{X}^2 \ddot{X}^2 - (\dot{X}\ddot{X})^2)^
\frac{1}{2}}{\dot{X}^2}  dt + \beta \int (\dot{X}^2)^\frac{1}{2} dt ,
\label{param}
\end{equation}
Pisarski showed, using perturbation theory, that at $\chi^{-1}=0$ the theory
is asymptotically free.  Therefore, the paths will become smoother as we go to
shorter distances, but longer distances are outside the validity of
perturbation theory.

Ambj\o rn et al.~\cite{ADJ87} studied the non-perturbative regime of this
theory but considered open endpoints. Using a lattice-like regularization and
a generalized central limit theorem, they concluded that there are actually no
other phase transitions than the $\chi^{-1}=0$ point (which it is expected
from asymptotic freedom). This implies that the correlation length must
increase for small $\chi^{-1}$.  A similar result was obtained in the
approximation of mean-field theory~\cite{AE87}.

The actual action we have simulated is a lattice transcription of the action
(\ref{param}) obtained by simply replacing the derivatives of the path
$\dot{X^\mu_i }$ by finite differences $X^\mu_{i+1}-X^\mu_i$ (and doing
analogously with second derivatives):
\begin{equation}
S=\beta\sum_{i=1}^{N} |X_i - X_{i-1}| + 2\chi\sum_{i=1}^{N}
\av{\sin \frac{\theta_i}{2}}.
\end{equation}
where $\theta_i$ is the angle between two adjacent segments.

The partition function of this theory is, then, defined as
\begin{equation}
Z=\int\prod^{N-1}_{i=1} d^3 X_i \; e^{- S}.
\end{equation}

It is easy to check that, like in the surface case, this action is invariant
under reparameterizations which may be thought as a kind of gauge symmetry.
This fact allows to fix the coupling $\beta=1$ obtaining, actually, a single
parameter action.

We have restricted ourselves to the case of closed paths, {\it i.e.\/} we have
require that the neighbor of the $ i=N $ point is $i=1$. We have also fixed
this point to avoid translational invariance. All these prescriptions are in
perfect agreement with the simulation of random surfaces.

{} From a geometrical point of view, the paths we have simulated consist in
closed loops of straight segments embedded in $\mbox{\QRC R}^3$, being this
topology preserved during the simulation.

\section{Numerical results}
We have performed numerical simulations for paths embedded in ${\QRC R}^D$
($D=3$) with different number of points: $N=25,100,200$ and for different
values of the curvature coupling $\chi$ from 0 to 10, keeping always
$\beta=1$.  The number of sweeps has been of the order of 2-4 millions for
each coupling.  The numerical computation of the partition function has been
done using a simple Metropolis algorithm~\cite{BGM84}.  We have not applied
histogram or re-summation techniques due to the large fluctuations and large
correlations experienced by this system.  The use of other methods such as
Hybrid Monte Carlo algorithms seems not to improve the results~\cite{BC97}.
Main results concerning the geometrical and thermodynamical behavior of the
paths are as follows:

\begin{enumerate}
  
\item {\bf Length.}  The measure of the mean length of the paths constitutes a
  good test of the numerical procedure.  It is easy to deduce that
  independently of the curvature $\chi$ value,
\begin{equation}
\ev{L}=-\frac{\partial \ln Z}{\partial\beta}=\frac{(N-1)D}{\beta}. 
\end{equation}
Results to compare with the analytical prediction are summarized in
table~\ref{tab:Lenghts}.  \tabLenghts

\item{\bf Curvature.} The mean total curvature of the path is given by
\begin{equation}
\ev{S_c} = \ev{\int k\, ds} =
                  -\frac{\partial \ln Z}{\partial\chi}.
\end{equation}

We have summarized in fig.~\ref{fig:Curvature} the results concerning the
curvature measurements.  Notice the decrease of the mean curvature as the
coupling $\chi$ grows.
The solid curve represents the limit $\chi\rightarrow\infty$ and $N
\rightarrow\infty$ when the path becomes a flat circle and the total curvature
is just $2\pi$.

\figCurvature

\figHausdorff

\item {\bf Hausdorff Dimension.}  The square gyration radius is a geometrical
  magnitude defined as
\begin{equation}
\ev{X^2}_c=\frac{1}{D}\pa{\ev{X^2}-\ev{\vphantom{X^2}X}^2}.
\end{equation}
For zero curvature coupling the mean value of the square gyration radius is
just~\cite{G84} 
\begin{equation}
\ev{X^2}_c= N\;\frac{(D+1)}{12\beta^2},
\label{eq:Hauss_1}
\end{equation}
in perfect agreement with our results. In the infinite curvature limit~
\cite{BC97}
\begin{equation}
\ev{X^2}_c=(N-1)^2\;\frac{D}{(2\pi\beta)^2}.
\end{equation} 
{} From the square gyration radius we can extract the Hausdorff dimension (see
fig.~\ref{fig:Hausdorff}).


{} From the analytical results~\cite{ADJ87} we know that in the continuous
limit random paths must be crumpled for all finite curvature couplings
(including the case $\chi=0$). This implies $d_H=2$ (i.e. the square gyration
radius scales as $N$). Rigid paths are expected only at the infinite curvature
coupling, and in this case, we expect $d_H=1$ (i.e. the square gyration radius
scales as $N^2$). %
The analysis of fig.~\ref{fig:Hausdorff} shows an unexpected behavior. For
$\chi=0$ the values of the Hausdorff dimension, for the different sizes of the
paths, are all higher than $d_H=2$. 
These values decrease when $N$ increases and they are in perfect agreement
with eq.~(\ref{eq:Hauss_1}), so in the infinite volume ($N\to\infty$) the value
of the Hausdorff dimension is $d_H=2$, as expected for Brownian paths.
However, the behavior of paths at non-zero
curvature coupling is rather difficult to connect with the above scenario.  We
observe that for each size $N$, the value of $d_H$ decreases as a function of
$\chi$ up to an almost constant value for large (but still finite) couplings.
For all the sizes studied, these limiting values are lower than $d_H=2$ and,
for large $N$, they approaches a value close to $d_H=1.5$.  The effects of
dealing with paths of finite size seems to be, then, stronger than naive
expectations.

We have drawn in fig.~\ref{fig:SS} simple snapshots of the paths for different
curvature couplings.  They show an apparent change of behavior from {\em
  crumpled} or Brownian paths to {\em rigid} paths.  For large curvature
terms, a closed path build of a finite number of straight segments seems to be
{\em frozen}.

\item{\bf Specific heat.}  In addition to this geometrical picture, we have
  measured also the heat capacity with respect to curvature, defined as
\begin{equation}
C_\chi = \frac{-\chi^2}{N} \;\; \frac{\partial^2}{\partial\chi^2}\ln Z,
\end{equation}
where $Z$ is the canonical partition function of the system.

\figSS

\figCv

Its behavior, shown in fig.~\ref{fig:Cv}, is really suggestive.  Remark, for
instance, the saturation for long paths.  The shape of this graph is very
similar to that corresponding to the self-avoiding fluid
membranes~\cite{KG92a}, for which the authors conclude the absence of a phase
transition.  Moreover it is also similar to the case of crystalline random
surfaces, where there is a true phase transition~\cite{BET94,WHE96}.  This
reflects the fact that a {\em crude} observation of a structure in the
specific heat graph is not relevant enough to characterize a phase transition.

\end{enumerate}

We should remark that a great difficulty in this analysis is the large
correlation found in the simulations, specially for large curvature couplings.
This is not only due to the poor efficiency of the method, but is mainly due
to the finite size effects of the system.

\section{Conclusions}
We have performed a numerical simulation of an ensemble of closed random paths
embedded in $\mbox{\QRC R}^3$ weighted with an action proportional to both,
the length of the path and the curvature.  Several geometrical and
thermodynamical magnitudes have been measured: Length, curvature, gyration
radius, Hausdorff dimension and specific heat.

This is a privileged framework, because from analytical results we know that
there is no true phase transition and that for a finite non-zero curvature
coupling these paths will belong to the same universality class than Brownian
random walks.  In addition, one has the possibility of perform a high
statistics simulation.  Nevertheless, in an actual numerical simulation, the
finite size effects and the large correlations in the system hide the expected
behavior.

In addition to the intrinsic interest of such a simulation --that completes
previous works on this field~\cite{PIS86,ADJ87}-- the experience acquired
warns us that one must be very careful when dealing with other situations
without firmly established analytical results.

\vspace{18pt}

\noindent {\bf Acknowledgments}

\noindent
Two of us (MB, JC) thank J. Ambj\o rn, D. Espriu, A. Pap and P. Suranyi for
useful comments and suggestions.  AJ is supported by a doctoral fellowship
from IVEI.  This work has been partially supported by research project
CICYT~AEN95/0882.

\end{document}